\documentstyle[]{article}
\font\cero=cmss10 scaled 1728
\font\uno=cmssbx10 scaled 1200
\begin{document}
\begin{flushleft}
{\cero Global symplectic potentials on the Witten covariant phase space
for bosonic extendons} \\[3em]
\end{flushleft}
{\sf R. Cartas-Fuentevilla}\\
{\it Instituto de F\'{\i}sica, Universidad Aut\'onoma de Puebla,
Apartado postal J-48 72570, Puebla Pue., M\'exico (rcartas@sirio.ifuap.buap.mx).}  \\[4em]

It is proved that the projections of the deformation vector field,
normal and tangential to the worldsheet manifold swept out by
Dirac-Nambu-Goto bosonic extendons propagating in a curved
background, play the role of {\it global} symplectic potentials on
the corresponding Witten covariant phase space. It is also proved
that the {\it presymplectic} structure obtained from such
potentials by direct exterior derivation, has not components
tangent to the action of the relevant diffeomorphisms group of the
theory. \\

\noindent Running title:  Global symplectic....\\

\noindent {\uno I. INTRODUCTION}
\vspace{1em}

The purpose of the present letter is to extend and remark our
recent analysis \cite{1} on the symplectic geometry of the Witten
covariant phase space for Dirac-Nambu-Goto(DNG) bosonic extendons
or p-branes. In that letter we have demonstrated, using a
covariant description of the deformation dynamics of such
extended objects, that the phase space ($Z$) is endowed with a
(pre)symplectic structure. Such a geometrical structure is
covariant in the strong sense of being expressed in terms of
ordinary tensor fields defined with respect to background
coordinate systems, and not referred to distinct worldsheet and
external background quantities (a remarkable virtue of the Carter
formalism for the deformation dynamics of extendons \cite{2,3}).
Furthermore, in our analysis \cite{1}, a gauge-fixing condition
on the deformation dynamics was imposed, specifically the so
called {\it orthogonal gauge}, which considers the deformations
orthogonal to the worldsheet as the only physically relevant, and
ignores the deformations tangent to the worldsheet, since can
always be identified with the action of a worldsheet
diffeomorphism. Another limitation of that analysis is that the
existence of degenerate directions on the covariant phase space ,
and associated with the gauge transformations of the extendon
theory, was not considered explicitly. Such gauge directions may
lead to a possible degeneracy of the presymplectic form, which in
turns, may affect the invertibility of such a differential form,
in order to construct the Poisson brackets, which are essential
for a possible quantum mechanical treatment of the theory.

Therefore, the purpose of the present analysis is twofold. First,
we shall demonstrate that in the general case where no
gauge-fixing condition on the deformation dynamics is imposed,
there exists an {\it exact} two-form on the phase space obtained
from certain one-forms by exterior derivation. Such a two-form
will be, in particular, {\it closed}, as required for constructing
a (pre)symplectic form on Z. It is shown that the one-forms
(called symplectic potentials), have a remarkably simple form in
terms of fundamental fields describing the deformation dynamics.
Additionally, it is also proved that the gauge directions of the
extendon theory under consideration, are eliminated in passing
from $Z$ to the reduced phase space (the space of solutions modulo
gauge transformations), obtaining finally a nondegenerate
geometrical structure for the theory.

At this point, it is opportune to review briefly the main results
of the letter \cite{1}, which will be useful in the present
analysis. Working in the orthogonal gauge,
\begin{equation}
     \eta^{\mu} {_{\nu}} \xi^{\nu} = 0,
\end{equation}
where $\eta^{\mu}{_{\nu}}$ is the {\it fundamental tensor} of the
worldsheet (that together with the complementary orthogonal
projection $\bot^{\mu} {_{\nu}}$ satisfy $g^{\mu\nu} =
\eta^{\mu\nu} + \bot^{\mu\nu}$, being $g_{\mu\nu}$ the background
metric), and $\xi^{\nu} \equiv \delta X^{\nu}$ the deformation
vector field of the worldsheet, the {\it symplectic} current that
is worldsheet covariantly conserved,
\begin{equation}
\widetilde{\cal J}^{\nu} = \xi^{\rho} \widetilde{\nabla}^{\nu}
     \xi_{\rho},   \qquad  \widetilde{\nabla}_{\nu} \widetilde{\cal J}^{\nu} = 0,
\end{equation}
permits to construct a two-form $\omega$ on the phase space $Z$
of DNG bosonic extendons, given by
\begin{equation}
     \omega \equiv \int_{\Sigma} \sqrt{- \gamma} \widetilde{\cal
J}^{\mu} d \widetilde{\Sigma}_{\mu},
\end{equation}
independent on the choice of $\Sigma$ (a spacelike section of the
worldsheet corresponding to a Cauchy $p$-surface for the
configuration of the extendon), and in particular Poincar\'{e}
invariant. $d \widetilde{\Sigma}_{\mu}$ is the surface measure
element of $\Sigma$, and is normal on $\Sigma$ and tangent to the
worldsheet, and  $\gamma$ is the determinant of the world sheet
metric. In \cite{1}, it is proved, by direct calculation of the
exterior derivative of Eq.\ (3), that $\omega$ is {\it
identically} closed on $Z$, $\delta\omega=0$, representing then a
(pre)symplectic structure on $Z$. However, in the general
situation where a gauge-fixing condition is no imposed, and Eq.\
(1) is no necessarily satisfied, the symplectic current has the
more complicated form \cite{1},
\begin{equation}
     \widetilde{\cal J}^{\nu} = (\eta^{\mu\nu} \bot_{\sigma\rho} +
2 \eta^{\nu} {_{[\sigma}} \eta_{\rho ]} {^{\mu}}) {\xi}^{\sigma}
\widetilde{\nabla}_{\mu} \xi^{\rho},
\end{equation}
and although is also worldsheet covariantly conserved \cite{1},
\begin{equation}
      \widetilde{\nabla}_{\nu} \widetilde{\cal J}^{\nu} = 0,
\end{equation}
it remains the question whether $\omega$ in (3), with
$\widetilde{\cal J}^{\nu}$ being of the general form (4), is
closed. Although one can follow exactly the same procedure used
in \cite{1} for answering this question, we shall face the
problem using a different strategy, which will have the virtue of
reveling a rich underlying structure of the phase space for DNG
extendons.

In next Section the explicit forms and properties of the
symplectic potentials on $Z$ are discussed. In Section III we
discuss the issue of degenerate directions and the invariance
properties of the relevant geometrical quantities on $Z$. We
conclude in Section IV with some remarks and prospects for the
future. \\[2em]

\noindent {\uno II. GLOBAL SYMPLECTIC POTENTIALS}
\vspace{1em}

As it is well known, on any manifold with symplectic structure
$\omega$, locally one can introduce coordinates $p_{i}$ and
$q_{i}$ such that $\omega = \delta (q_{i} \delta p_{i})$, where
$q_{i} \delta p_{i}$ is an one-form called the {\it canonical
symplectic potential}. In general, such a potential exists only
locally and is not unique.

In this section we shall demonstrate that, without invoking local
canonical coordinates (which is in fact the spirit of the present
covariant description of the phase space), there exists a {\it
global} symplectic potential on $Z$, which generates by direct
exterior derivation the symplectic current (4). The demonstration
is very simple, and consists in to rewrite $\widetilde{\cal
J}^{\nu}$ in a way such that we can identify directly the
corresponding symplectic potential.

Hence, using the definition $\widetilde{\nabla}_{\mu} \equiv
\eta_{\mu} {^{\nu}} \nabla_{\nu}$, the property $\eta^{\mu}
{_{\rho}} \eta^{\rho} {_{\nu}} = \eta^{\mu} {_{\nu}}$, and the
decomposition $\bot_{\sigma\rho} = g_{\sigma\rho} -
\eta_{\sigma\rho}$, we can rewrite $\widetilde{\cal J}^{\nu}$ in
(4), after suitably grouping, as
\begin{equation}
     \widetilde{\cal J}^{\nu} = \eta^{\nu\alpha} [\xi_{\sigma}
\nabla_{\alpha} \xi^{\sigma} - \eta_{\sigma} {^{\rho}}
\xi^{\sigma} (\nabla_{\alpha} \xi_{\rho} +  \nabla_{\rho}
\xi_{\alpha}) + \eta_{\rho}{^{\sigma}} \xi_{\alpha}
\nabla_{\sigma} \xi^{\rho}];
\end{equation}
the first terms can be identified directly in terms of variations
of fundamental geometrical quantities, considering that $\delta
\xi_{\mu} =-\xi^{\nu}\nabla_{\mu} \xi_{\nu}$ (see Eq.\ (23) in
\cite{1}), and that  $\delta \eta^{\mu\nu} = - \eta^{\mu\rho}
\eta^{\nu\sigma} \delta g_{\rho\sigma}$ :
\begin{equation}
     \eta^{\nu\alpha} \xi_{\sigma} \nabla_{\alpha} \xi^{\sigma} =
- \eta^{\nu\alpha} \delta \xi_{\alpha}, \qquad \xi^{\sigma}
\eta^{\nu\alpha} \eta_{\sigma} {^{\rho}} (\nabla_{\alpha}
\xi_{\rho}
     +  \nabla_{\rho} \xi_{\alpha}) = - \xi_{\sigma} \delta
\eta^{\nu\sigma},
\end{equation}
and the last term can be rewritten as
\begin{equation}
     \eta^{\nu\alpha} \xi_{\alpha} \eta_{\rho} {^{\sigma}}
\nabla_{\sigma} \xi^{\rho} = \frac{1}{2} (\eta^{\nu\alpha}
\xi_{\alpha}) \eta^{\sigma\rho} [\nabla_{\sigma} \xi_{\rho} +
\nabla_{\rho} \xi_{\sigma}] = \eta^{\nu\alpha} \xi_{\alpha}
\frac{\delta \sqrt{-\gamma}}{\sqrt{-\gamma}},
\end{equation}
where the last equality follows from the formula
$\delta\sqrt{-\gamma}=\sqrt{-\gamma}\eta^{\mu\nu}\delta
g_{\mu\nu}$ \cite{3}. Therefore, from Eqs.\ (6), (7), and (8) we
have that,
\begin{equation}
     \sqrt{-\gamma} \widetilde{\cal J}^{\nu} = \sqrt{-\gamma}
(\eta^{\alpha\nu} \delta \xi_{\alpha} +  \delta \eta^{\alpha\nu}
\xi_{\alpha}) + (\eta^{\alpha\nu} \xi_{\alpha}) \delta
\sqrt{-\gamma} = \delta (-\sqrt{-\gamma} \eta^{\alpha\nu}
\xi_{\alpha}),
\end{equation}
where we have considered the Leibniz rule  for $\delta$, and the
fact that $\eta^{\alpha\nu} \xi_{\alpha}$ and $\delta
\sqrt{-\gamma}$ correspond to one-forms on $Z$ and thus are
anticommutating objects \cite{1}. Therefore, Eq.\ (9) shows that
the smooth one-form $\omega^{\nu} \equiv - \sqrt{-\gamma}
\eta^{\alpha\nu} \xi_{\alpha}$ is a global symplectic potential
density on $Z$ for DNG extendons. Hence, Eq.\ (9) permits to write
out $\omega$ (see paragraph after Eq.\ (5)) as
\begin{equation}
     \omega = \int_{\Sigma} \delta \omega^{\nu}
d\widetilde{\Sigma}_{\nu},
\end{equation}
which shows that $\omega$ is an {\it exact} differential form, and
in particular, an identically closed two-form on $Z$, since
$\delta$ is nilpotent,
\begin{equation}
     \delta \omega = 0,
\end{equation}
as required for constructing a (pre)symplectic structure on the
phase space, in this general case where no gauge-fixing condition
(Eq.\ (1)) is imposed. Note that Eq.\ (5) guarantees that
$\omega$ in (10) is independent on the choice of $\Sigma$ and, in
particular, is Poincar\'{e} invariant.

It is worth pointing out some properties of the symplectic
potential $\omega^{\nu}$. First, $\omega^{\nu}$ is not unique, and
the form of the ambiguity is easily determined from Eq.\ (9), and
using again the nilpotency of $\delta$: it is defined up to the
exterior derivative of any (background) vector field, say
$\lambda^{\nu}$, since
\begin{equation}
     \sqrt{-\gamma} \widetilde{\cal J}^{\nu} = \delta (\omega^{\nu} +
\delta \lambda^{\nu}).
\end{equation}
Thus, one can think of $\omega^{\nu}$ and $\omega^{\nu} + \delta
\lambda^{\nu}$ as gauge fields on the phase space, which
correspond to the same field strength $\omega$, the only
physically meaningful geometrical structure on $Z$. Furthermore,
using again the decomposition $\eta^{\nu}{_{\alpha}} = g^{\nu}
{_{\alpha}} - \bot^{\nu} {_{\alpha}}$, we can rewrite
$\omega^{\nu}$ as
\begin{equation}
     \omega^{\nu} = \sqrt{-\gamma} \bot^{\nu} {_{\alpha}}
\xi^{\alpha} - \sqrt{-\gamma} \xi^{\nu} = \sqrt{-\gamma}
\bot^{\nu} {_{\alpha}} \xi^{\alpha} - \delta (\sqrt{-\gamma}
X^{\nu}),
\end{equation}
where we have considered that $\delta\sqrt{-\gamma}=0$,
corresponds to the first order action variation \cite{1}, and the
definition $\xi^{\nu} \equiv \delta X^{\nu}$ for rewriting the
second term as the exterior derivative of the zero-form
$\sqrt{-\gamma} X^{\nu}$, which can now be identified, in
particular, with the above vector field $\lambda^{\nu}$.
Therefore, from Eqs.\ (9), and (13) we can see that both the
tangential projection $\eta^{\nu} {_{\alpha}} \xi^{\alpha}$ to
the world-sheet of the infinitesimal deformation $\xi^{\nu}$
(identified with the action of a dynamical world-sheet
infinitesimal diffeomorphism) and the orthogonal projection
$\bot^{\nu} {_{\alpha}} \xi^{\alpha}$ (identified with the
physically observable measure of the deformation), play the role
of global symplectic potentials on the phase space $Z$, with the
background coordinate field $(\sqrt{-\gamma}) X^{\nu}$ playing
the role of the generating function of a transformation from the
(global) variables $(\eta^{\alpha} {_{\nu}}, X^{\nu})$ to
$(\bot^{\alpha} {_{\nu}}, X^{\nu})$, and conversely, in according
to Eq.\ (13). Furthermore, note that specifically in the case of
the tangential projection, which can be ignored since is not
physically observable as a deformation, its exterior derivative
on the phase space is, in accordance with Eq.\ (10), physically
meaningful. Hence, a quantity that is ``pure gauge" in the
conventional deformation scheme, generates by direct exterior
derivation a physically relevant geometrical structure $\omega$
on the (nonconventional) phase space $Z$. \\[2em]

\noindent {\uno III. THE SYMPLECTIC STRUCTURE ON THE REDUCED PHASE
SPACE} \vspace{1em}

Another aspect well known of the covariant phase space formulation
is that, by defining the phase space $Z$ as the space of solutions
dynamically allowed by the classical equations of motion, there
exist naturally degenerate directions, which will be revealed in
the degeneracy of the geometrical structure $\omega$. More
specifically, the motions along such directions correspond to the
gauge transformations of the theory. In the present case of DNG
extendons embedded in a curved spacetime (such as any theory
formulated in terms of a spacetime manifold and tensor fields
defined on it), the degenerate directions will correspond to
spacetime infinitesimal diffeomorphisms, and particularly to world
sheet infinitesimal reparametrizations. In this sense, the
geometrical structure $\omega$ previously constructed is actually
a {\it presymplectic} form. Therefore, our purpose in this section
is to demonstrate that $\omega$ has not effectively components
along such gauge directions. The argument is again very simple and
consists essentially in proving the invariance of the deformation
vector field $\xi^{\nu}$ under spacetime infinitesimal
diffeomorphisms,
\begin{equation}
     X^{\mu} \rightarrow  X^{\mu} + \delta X^{\mu}.
\end{equation}
Note that physically $\xi^{\nu} \equiv \delta X^{\nu}$ represents
deformations of the worldsheet geometry, whereas $\delta X^{\mu}$
in (14), a diffeomorphism. Under such a transformation $\xi^{\nu}$
changes by
\begin{eqnarray}
     \xi^{\mu}  \!\! & \rightarrow  & \!\! \delta (X^{\mu} +
\delta X^{\mu}) \nonumber \\
\!\! & & \!\! = \xi^{\mu} + \delta^{2} X^{\mu},
\end{eqnarray}
we mean, by a second order term (as it must!), which is negligible
in a first order deformation scheme. However, $\delta^{2} X^{\mu}$
vanishes strictly on the phase space, in virtue of the nilpotency
property, and therefore the deformation $\xi^{\nu}$, an one-form
on $Z$, has not components along the gauge directions (as
awaited!, since $\xi^{\nu}$ is ``the physically observable
measure" of the deformations of the worldsheet geometry). As a
consequence, any object on $Z$ that depends on $\xi^{\nu}$, also
will have vanishing components along such directions. Hence, if
$Z$ is the space of solutions of the DNG extendon dynamics, and
$\widehat{Z}$ is the space of solutions modulo gauge
transformations (we mean, $\widehat{Z}$ is the {\it quotient}
space $Z$/$G$, or reduce phase space, with $G$ being the group of
diffeomorphisms), then $\widetilde{\cal J}^{\nu}$ and $\omega$
have vanishing components along the $G$ orbits, and specifically
$\omega$ will be a nondegenerate two-form on $\widehat{Z}$. Note
that, among all degenerate directions  on $Z$, there will be some
directions associated to worldsheet reparametrizations, and then
the diffeomorphism in Eq.\ (14) can take, in particular, the form
of a worldsheet diffeomorphism,
\begin{equation}
     \delta X^{\mu} = \epsilon^{\alpha} \partial_{\alpha} X^{\mu},
\end{equation}
where $\epsilon^{\alpha}$ is a infinitesimal shift in the
worldsheet coordinates, and $\partial_{\alpha}$ denotes derivation
with respect to them. Then $\widetilde{\cal J}^{\nu}$ and
$\omega$ are, in particular, invariant under (16).

There exists another form for demonstrating the invariance of
$\widetilde{\cal J}^{\nu}$ and $\omega$ under diffeomorphisms, and
consists in realizing the role of the field $\lambda^{\nu}$ in
(12). As we have seen, $\omega^{\nu}$ corresponds to projections
of the variation of the background coordinate field $X^{\nu}$,
and then $\delta \lambda^{\nu}$ in (12) can be understood, in
particular, as an infinitesimal diffeomorphism, $\delta
\lambda^{\nu} = \delta X^{\nu}$, and thus the invariance of
$\widetilde{\cal J}^{\nu}$ and $\omega$ is guaranteed from (12).
Thus, the indeterminacy directions of the symplectic potential can
correspond, in particular, to the gauge directions of the
theory.  Therefore, $\omega$ is finally our covariant and gauge
invariant description of the canonical formalism for DNG bosonic
extendons. \\ [2em]

\noindent {\uno IV. REMARKS AND PROSPECTS} \vspace{1em}

As mentioned, locally $\omega = \delta q_{i} \delta p_{i}$, and
can be described in terms of a canonical symplectic potential in
the forms $\omega = \delta (q_{i} \delta p_{i})$ or $\omega =
\delta (-p_{i} \delta q_{i})$, with $p_{i}$ and $q_{i}$ being the
conventional canonically conjugate variables. Similarly, in the
present covariant formulation, $\widetilde{\cal J}^{\nu}$ (and
hence $\omega$) can be written as
\begin{equation}
     \widetilde{\cal J}^{\nu} = \delta (\eta_{\alpha} {^{\nu}}
\xi^{\alpha}) = \delta (\eta_{\alpha} {^{\nu}} \delta X^{\alpha})
= \delta (- X^{\alpha} \delta \eta_{\alpha} {^{\nu}}),
\end{equation}
in virtue again of the Leibniz rule and the nilpotency property,
and similarly for the pair $(\bot^{\nu} { _{\alpha}},
X^{\alpha})$. Elucidating, we can think of $(\eta_{\alpha}
{^{\nu}}, X^{\alpha})$ as {\it conjugate} variables in this
covariant description of the phase space for DNG extendons. This
question, and other relevant aspects of the transition between the
classical and quantum domains, require of course of a deeper
research, and will be the subject of forthcoming communications.

 Although we have studied only the bosonic case, physically
more interesting theories such as {\it superextendons}, will also
be a problem for the future.

In a recent paper \cite{4} we have obtained, using a different
scheme for the deformations dynamics of extended objects, similar
results to those presented in \cite{1}; however, that scheme is
{\it weakly} covariant. \\ [2em]

\begin{center}
{\uno ACKNOWLEDGMENTS}
\end{center}
\vspace{1em}

This work was supported by CONACyT and the Sistema Nacional de
Investigadores (M\'exico). The author wants to thank H. Garcia
Compean for drawing my interest to the study of extendons, and Dr.
G. F. Torres del Castillo for discussions.
\\[2em]


\begin{thebibliography}{}
\setlength{\itemsep}{-.50em}
\bibitem{1} R. Cartas-Fuentevilla, {\it Identically closed two-form for
covariant phase space quantization of Dirac-Nambu-Goto p-branes in
a curved spacetime}, to be published, Phys.\ Lett.\ B, (2002).
\bibitem{2} B. Carter, Phys.\ Rev.\ D {\bf 48}, 4835 (1993).
\bibitem{3} B. Carter, 1997 {\it Brane dynamics for treatment of cosmic
strings and vortons}, in {\it Recent Developments in Gravitation
and Mathematics, Proc. 2nd Mexican School on Gravitation and
Mathematical Physics (Tlaxcala, 1996)}
(http://kaluza.physik.uni-konstanz.de/2MS) ed. A. Garcia, C.
Lammerzahl, A. Macias and D. Nu\~{n}ez (Konstanz: Science
Network).
\bibitem{4} R. Cartas-Fuentevilla, {\it Towards a covariant canonical
quantization for closed topological defects without bondaries},
submitted to Phys. Lett. B, (2002).

\end{thebibliography}
\end{document}